\documentclass[%
 reprint,
superscriptaddress,
 amsmath,amssymb,
aps,
prl,
]{revtex4-2}

\usepackage{graphicx}
\usepackage{dcolumn}
\usepackage{bm}
\usepackage{upgreek}
\usepackage{esvect}
\usepackage{siunitx}

\begin{document}

\title{Electronic structure of the alternating monolayer-trilayer phase of La$_3$Ni$_2$O$_7$}

\author{Sebastien Abadi}
 \altaffiliation[]{These authors contributed equally to this work.}
 \affiliation{%
 Stanford Institute for Materials and Energy Sciences, SLAC National Accelerator Laboratory, 2575 Sand Hill Road, Menlo Park, California 94025, USA
 }%
 \affiliation{%
 Geballe Laboratory for Advanced Materials, Department of Physics and Applied Physics, Stanford University, Stanford, California 94305, USA \\
 }
 \affiliation{%
 Department of Physics, Stanford University, Stanford, CA, 94305, USA}

\author{Ke-Jun Xu}
 \altaffiliation[]{These authors contributed equally to this work.}
 \affiliation{%
 Stanford Institute for Materials and Energy Sciences, SLAC National Accelerator Laboratory, 2575 Sand Hill Road, Menlo Park, California 94025, USA
 }%
 \affiliation{%
 Geballe Laboratory for Advanced Materials, Department of Physics and Applied Physics, Stanford University, Stanford, California 94305, USA \\
 }
 \affiliation{%
 Department of Applied Physics, Stanford University, Stanford, CA, 94305, USA}

\author{Eder G. Lomeli}
 \altaffiliation[]{These authors contributed equally to this work.}
  \affiliation{%
 Stanford Institute for Materials and Energy Sciences, SLAC National Accelerator Laboratory, 2575 Sand Hill Road, Menlo Park, California 94025, USA
 }%
 \affiliation{Department of Materials Science and Engineering, Stanford University, Stanford, CA, 94305, USA}

\author{Pascal Puphal}
  \affiliation{Max Planck Institute for Solid State Research, Heisenbergstraße 1, D-70569 Stuttgart, Germany}

\author{Masahiko Isobe}
 \affiliation{Max Planck Institute for Solid State Research, Heisenbergstraße 1, D-70569 Stuttgart, Germany}

\author{Yong Zhong}
 \affiliation{%
 Stanford Institute for Materials and Energy Sciences, SLAC National Accelerator Laboratory, 2575 Sand Hill Road, Menlo Park, California 94025, USA
 }%
 \affiliation{%
 Geballe Laboratory for Advanced Materials, Department of Physics and Applied Physics, Stanford University, Stanford, California 94305, USA \\
 }
 \affiliation{%
 Department of Applied Physics, Stanford University, Stanford, CA, 94305, USA}

\author{Alexei V. Fedorov}
 \affiliation{Advanced Light Source, Lawrence Berkeley National Laboratory, Berkeley, CA, 94720, USA}

\author{Sung-Kwan Mo}
 \affiliation{Advanced Light Source, Lawrence Berkeley National Laboratory, Berkeley, CA, 94720, USA}

\author{Makoto Hashimoto}
 \affiliation{Stanford Synchrotron Radiation Lightsource, SLAC National Accelerator Laboratory, Menlo Park, CA, 94305, USA}

\author{Dong-Hui Lu}
 \affiliation{Stanford Synchrotron Radiation Lightsource, SLAC National Accelerator Laboratory, Menlo Park, CA, 94305, USA}

\author{Brian Moritz}
 \affiliation{%
 Stanford Institute for Materials and Energy Sciences, SLAC National Accelerator Laboratory, 2575 Sand Hill Road, Menlo Park, California 94025, USA
 }%

\author{Bernhard Keimer}
 \affiliation{Max Planck Institute for Solid State Research, Heisenbergstraße 1, D-70569 Stuttgart, Germany}

\author{Thomas P. Devereaux}
\email{tpd@slac.stanford.edu}
  \affiliation{%
 Stanford Institute for Materials and Energy Sciences, SLAC National Accelerator Laboratory, 2575 Sand Hill Road, Menlo Park, California 94025, USA
 }%
 \affiliation{Department of Materials Science and Engineering, Stanford University, Stanford, CA, 94305, USA}

\author{Matthias Hepting}
\email{hepting@fkf.mpg.de}
 \affiliation{Max Planck Institute for Solid State Research, Heisenbergstraße 1, D-70569 Stuttgart, Germany}

\author{Zhi-Xun Shen}
\email{zxshen@stanford.edu}
 \affiliation{%
 Stanford Institute for Materials and Energy Sciences, SLAC National Accelerator Laboratory, 2575 Sand Hill Road, Menlo Park, California 94025, USA
 }%
 \affiliation{%
 Geballe Laboratory for Advanced Materials, Department of Physics and Applied Physics, Stanford University, Stanford, California 94305, USA \\
 }
 \affiliation{%
 Department of Physics, Stanford University, Stanford, CA, 94305, USA}
  \affiliation{%
 Department of Applied Physics, Stanford University, Stanford, CA, 94305, USA}
 
\date{\today}

\begin{abstract}

Recent studies of La$_3$Ni$_2$O$_7$ have identified a bilayer (2222) structure and an unexpected alternating monolayer-trilayer (1313) structure, both of which feature signatures of superconductivity near 80~K under high pressures. Using angle-resolved photoemission spectroscopy, we measure the electronic structure of 1313 samples. In contrast to the previously studied 2222 structure, we find that the 1313 structure hosts a flat band with a markedly different binding energy, as well as an additional electron pocket and band splittings. By comparison to local-density approximation calculations, we find renormalizations of the Ni-$d_{z^2}$ and Ni-$d_{x^2-y^2}$ derived bands to be about 5 to 7 and about 4 respectively, suggesting strong correlation effects. These results reveal important differences in the electronic structure brought about by the distinct structural motifs with the same stoichiometry. Such differences may be relevant to the putative high temperature superconductivity.

\end{abstract}

\maketitle

{\it Introduction}.\textemdash
The nickelates are close cousins to the cuprate family of high transition temperature ($T_c$) superconductors~\cite{keimer_quantum_2015}, sharing similar planar structural motifs, as well as correlated 3$d$ electron physics~\cite{greenblatt_ruddlesden-popper_1997, rodriguez-carvajal_neutron_1991,torrance_systematic_1992}. As a result, the nickelates have been a subject of study for decades~\cite{greenblatt_ruddlesden-popper_1997, rodriguez-carvajal_neutron_1991,torrance_systematic_1992}, including La$_3$Ni$_2$O$_7$ specifically, which was first synthesized as a polycrystalline powder in 1994~\cite{zhang_synthesis_1994}. In 2019, a breakthrough occurred with the first detection of superconductivity in a nickel-based oxide, with a $T_c$ of about 20~K achieved in the thin-film, infinite-layer nickelate Nd$_{0.8}$Sr$_{0.2}$NiO$_2$~\cite{li_superconductivity_2019}. Through the application of high pressure, the record $T_c$ in the infinite layer compounds was later increased up to 31~K~\cite{wang_pressure-induced_2022}.

The recent report of superconductivity approaching 80~K in pressurized single-crystal La$_3$Ni$_2$O$_7$~\cite{sun_signatures_2023} is a step beyond that, bringing nickelates into the realm of high temperature superconductivity. Consequently, this report has been met with tremendous interest, including considerable theoretical discussions of the possible underlying superconducting mechanism~\cite{luo_bilayer_2023,
 zhang_electronic_2023,
 yang_possible_2023,
 lechermann_electronic_2023,
 gu_effective_2023,
 shen1_effective_2023,
 sakakibara_possible_2024,
 shilenko_correlated_2023,
 lu_interlayer-coupling-driven_2024,
 liao_electron_2023,
 qu_bilayer_2024,
 yang_interlayer_2023,
 tian_correlation_2024,
 wu_superexchange_2024,
 liu_sifmmodepmelsetextpmfi-wave_2023,
 huang_impurity_2023,
 kun_jiang_high-temperature_2024,
 cao_flat_2024,
 qin_high-t_c_2023,
 chen_critical_2023,
 christiansson_correlated_2023,
 oh_type-ii_2023}. The initial crystal structure reported to be responsible for this high $T_c$ superconductivity is a bilayer (2222) perovskite structure consisting of two adjacent Ni-O layers separated from other Ni-O layers by La-O layers. However, since this discovery, several complications have arisen. For one, reports have been mixed on whether the superconductivity observed in La$_3$Ni$_2$O$_7$ is filamentary in nature~\cite{sun_signatures_2023, zhou_evidence_2024, li_pressure-driven_2024}, which alongside inconsistencies in $T_c$ and the presence of residual resistance in many samples~\cite{wang_pressure-induced_2024,
      zhang_high-temperature_2024,
      jun_hou1_emergence_2023,
      sui_electronic_2024,
      zhang_high-temperature_2024}, 
suggests that a minority phase may be responsible for the observed superconductivity. There have also been reports of superconductivity emerging under pressure in crystals exhibiting a majority phase of distinct structures: in the trilayer La$_4$Ni$_3$O$_{10}$ with a $T_c$ as high as 30~K~\cite{ li_signature_2024, zhang_superconductivity_2024, zhu_superconductivity_2024}, 
and in the alternating monolayer-trilayer (1313) structure La$_3$Ni$_2$O$_7$~\cite{puphal_unconventional_2023, chen_polymorphism_2024, wang_long-range_2024} 
with a $T_c$ also nearing 80~K~\cite{puphal_unconventional_2023}. These results lead to the important question: Which phase is truly responsible for the observed superconducting signatures in resistivity?

\begin{figure*}[t!]
\includegraphics[width=1\textwidth]{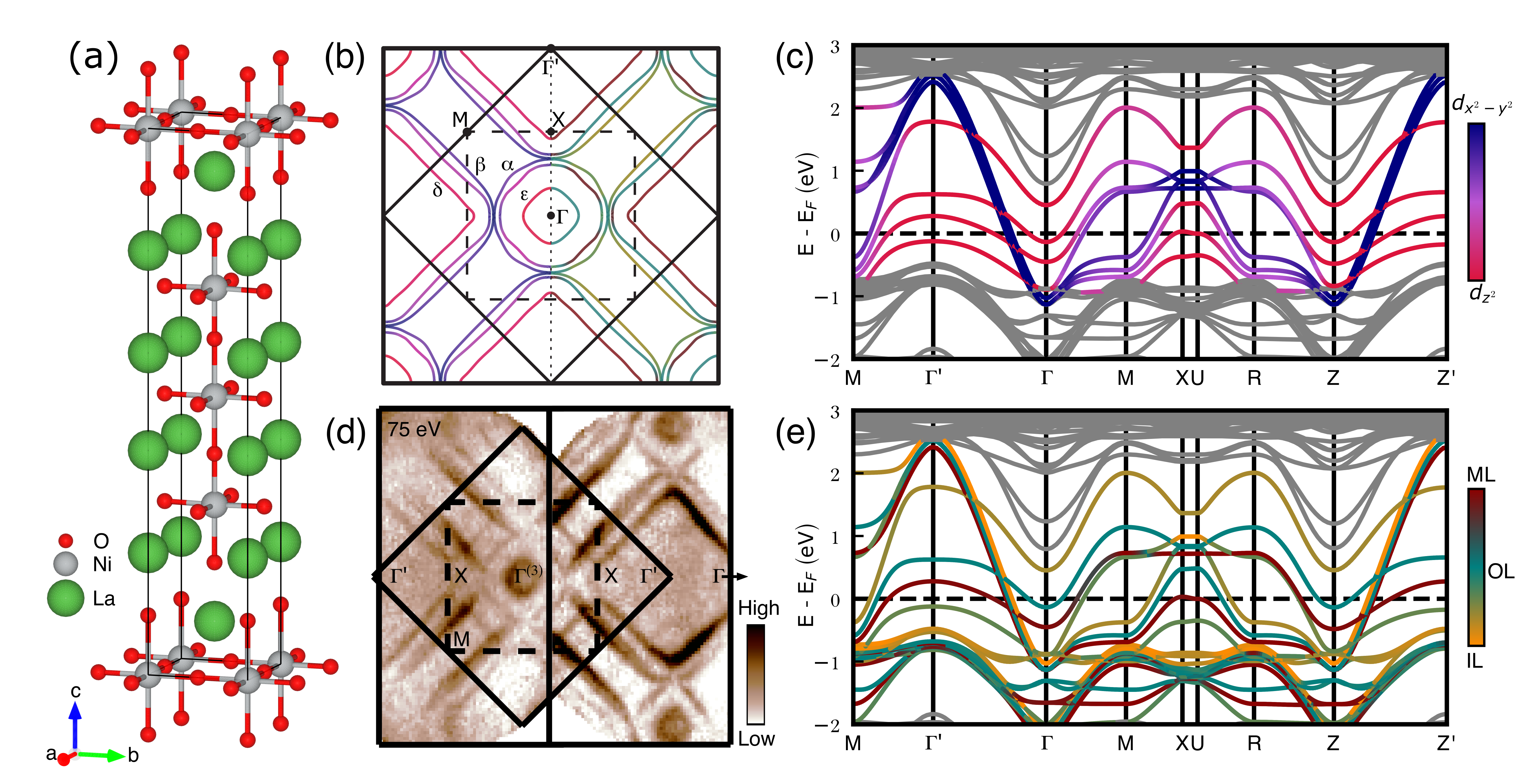}
\caption{Crystal structure, calculated and measured Fermi surface maps, and calculated band structure along high symmetry cuts. (a) Schematic of the 1313 crystal structure of La$_3$Ni$_2$O$_7$. The structure shown here is the unbuckled structure (space group $P4/mmm$) used in the accompanying LDA calculations. (b) Calculated Fermi surface map, where the color scheme on the left indicates relative Ni-$d_{x^2-y^2}$ versus Ni-$d_{z^2}$ orbital contributions, while the color scheme on the right indicates relative contributions of Ni in the monolayer (ML), and in the outer and inner layers of the trilayer (OL, IL respectively). (c),(e) Calculated band structures showing relative Ni-$d_{x^2-y^2}$ and Ni-$d_{z^2}$ orbital contributions and relative ML, OL, and IL contributions respectively. Note that the symmetry labels used in (b)-(c),(e) are of the reconstructed Brillouin zone (dashed lines in (b)), but calculations were performed using the unfolded zone (solid lines in (b)). (d) Experimentally determined Fermi surface map measured at 7 K with a photon energy of 75~eV. The left and right halves are integrated about E$_F$ in $\pm$35~mev and $\pm$15~meV windows respectively. The reconstructed Brillouin zone is indicated by black dashed lines, while the original zone is indicated by solid black lines. The left half of the map was measured centered at X between the second and fourth reconstructed Brillouin zones, while the right half was centered in the second reconstructed zone around $\Gamma'$.  
}
\end{figure*}

To help elucidate the above question, we use angle-resolved photoemission spectroscopy (ARPES) to investigate the electronic structure of La$_3$Ni$_2$O$_7$ crystals with the 1313 structure, and directly compare it to that of the 2222 structure~\cite{yang_orbital-dependent_2024, du2024correlated} and that of the trilayer La$_4$Ni$_3$O$_{10}$ structure~\cite{li_fermiology_2017, zhang_superconductivity_2024, du2024correlated}. We find an additional electron pocket consisting of Ni-$d_{z^2}$ orbital character and split flat bands near E$_F$, which may be relevant to superconductivity. The flat band originating from the trilayer sections of the 1313 structure lies within 20~mev below the Fermi energy (E$_F$), while the flat band originating from the monolayer sections passes above E$_F$. In a testament to the high data quality, we additionally resolve multilayer splitting effects in the large cuprate-like pockets centered at the Brillouin zone corners, which were predicted for
trilayer blocks of Ni-O planes, but were never experimentally observed thus far. By examining the dispersions of the bands near E$_F$, we determine strong renormalizations of about 4 for the Ni-d$_{x^2-y^2}$ derived bands and about 5 to 7 for the Ni-d$_{z^2}$ derived bands. Our results set the basis for further experimental and theoretical studies of the unusual 1313 structure in layered perovskite nickelates.

{\it Calculated Fermi surface and band structure}.\textemdash For crystals with the 1313 stacking configuration at ambient pressure, crystal structures with unbuckled and buckled Ni-O octahedra have both been reported ~\cite{puphal_unconventional_2023, chen_polymorphism_2024, wang_long-range_2024}, where the buckling is similar to what is observed in the ambient pressure 2222~\cite{sun_signatures_2023, ling_neutron_2000} and trilayer~\cite{li_fermiology_2017, ling_neutron_2000} structures. 
In particular, annealing the as-grown 1313 crystals converts the ambient pressure structure from unbuckled to buckled~\cite{puphal_unconventional_2023}. Signatures of superconductivity in the 1313 phase have only been reported in the annealed crystals at high pressure, where the structure reverts to unbuckled~\cite{puphal_unconventional_2023}. The present ARPES work was performed on as-grown crystals, with a bulk unbuckled structure [Fig. 1(a)]. These as-grown crystals at ambient pressure may serve as a reference for the high pressure unbuckled phase in the annealed crystals.

Figure 1(b)-(c),(e) shows the Fermi surface and band structure as found by local-density approximation (LDA) calculations for the unbuckled crystal structure [Fig. 1(a)] in space group $P4/mmm$~\cite{wang_long-range_2024}, as well as the relative orbital and layer contributions to those bands (see Supplementary Materials for calculation details). Six bands are predicted to cross the Fermi surface. In addition to the $\alpha$ and $\beta$ bands with Ni-$d_{x^2-y^2}$ orbital character observed in the 2222 structure~\cite{yang_orbital-dependent_2024}, a small electron pocket (labeled as $\varepsilon$) with Ni-d$_{z^2}$ orbital character is predicted around $\Gamma$. Also, a flat band that approaches E$_F$ near the $\Gamma'$ point, previously observed in the 2222~\cite{yang_orbital-dependent_2024} and trilayer~\cite{li_fermiology_2017} compounds, is predicted to split into two bands in the 1313 compound: one of primarily monolayer Ni-d$_{z^2}$
orbital character (labeled as $\delta$), and the other of trilayer Ni-d$_{z^2}$ orbital character (labelled as $\gamma$). While the $\gamma$ band is predicted to remain below E$_F$ as in the 2222~\cite{yang_orbital-dependent_2024} and trilayer~\cite{li_fermiology_2017} cases, the $\delta$ band is predicted to cross E$_F$.   Likewise, the $\alpha$ and $\beta$ bands are expected to exhibit stronger multilayer splitting than what is predicted for the 2222 structure~\cite{yang_orbital-dependent_2024, sun_signatures_2023}.

\begin{figure*}[t!]
\includegraphics[width=0.75\textwidth]{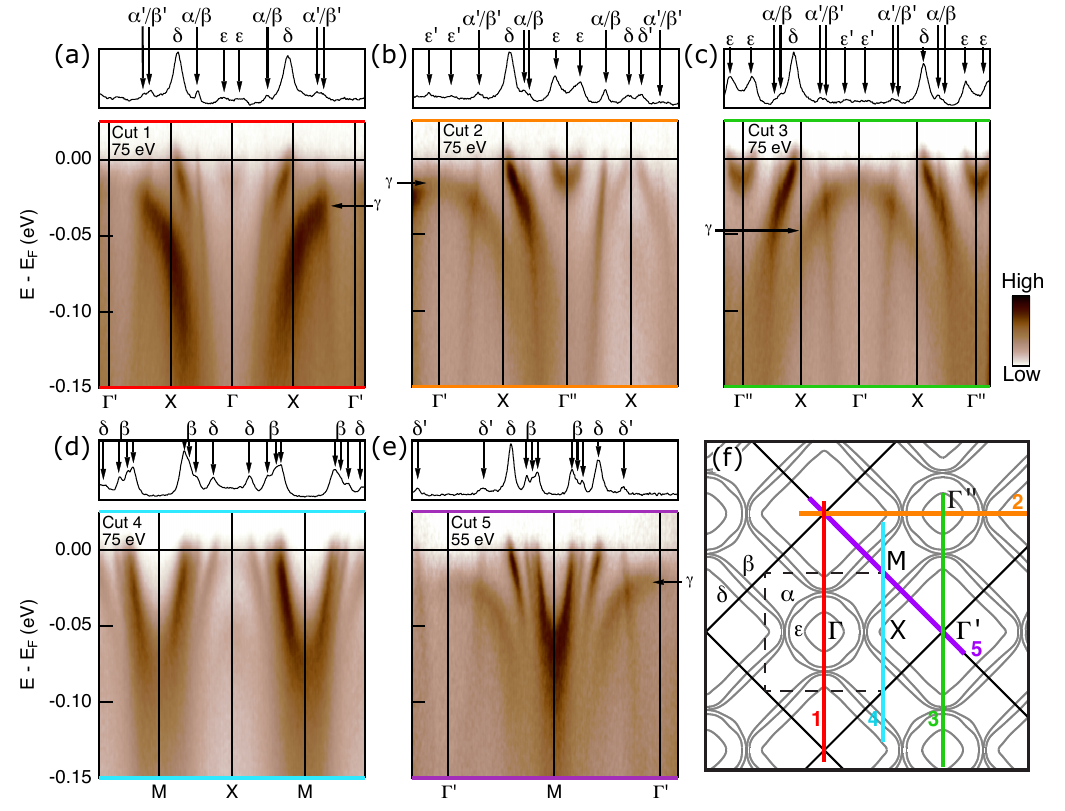}
\caption{Measured high symmetry cuts. (a)-(e) Band structure as measured along high symmetry cuts at 7 K, with photon energy of 75~eV for cuts (a)-(d) and 55~eV for cut (e). (f) Cut orientations. Above each cut, an MDC of that cut near E$_F$ is shown, and peaks corresponding to the bands crossing E$_F$ in the cut are labelled. In cuts (a)-(c), note that the $\alpha$ bands and $\beta$ bands are all nearly degenerate, and so are labelled as $\alpha/\beta$. Apostrophes indicate reconstructed bands.}
\end{figure*}

Band structure calculations were also performed using the buckled structure (space group $Fmmm$) as reported for the annealed 1313 crystals at 0.7~GPa~\cite{puphal_unconventional_2023} 
(see Supplementary Materials, Fig. S1)
. The enlarged unit cell in these calculations feature a ($\sqrt{2} \times \sqrt{2}$) reconstruction of the Brillouin zone, which is consistent with the experimental observation of folded $\alpha, \beta, \delta$, and $\varepsilon$ bands [Fig. 1(d), 2(a)-(c),(e)]. The observation of the ($\sqrt{2} \times \sqrt{2}$) reconstruction in experiment suggests that the bulk unbuckled as-grown crystals may experience buckling in the surface layers similar to the buckling present in the bulk of the ambient pressure annealed samples. For other features of the Fermi surface and band structure, calculations performed using the unbuckled and buckled structures give overall similar features, but the absence of folding in the unbuckled structure calculations makes it a clearer platform for comparison to the experimental results presented below.

{\it Experimental Fermi surface maps and band structure}.\textemdash Figure 1(d) shows an experimental Fermi surface map measured at 7~K using 75~eV photons. The left and right sections are integrated within $\pm$35~meV and $\pm$15~meV of E$_F$ respectively. Potentially due to surface reconstruction effects, such as Ni-O buckling in the topmost layers, the Brillouin zone is folded by a ($\sqrt{2} \times \sqrt{2}$) reconstruction.  The reconstructed Brillouin zone boundaries are indicated by dashed lines, while the original zone boundaries are indicated by solid lines. The left half of the map in Fig. 1(d) was measured centered at X between $\Gamma'$ and $\Gamma^{(3)}$ (at $\left(\frac{3\pi}{2},\ \frac{3\pi}{2}\right)$ of the original Brillouin zone), while the right half of the map was measured centered at $\Gamma'$ in the second reconstructed zone. In these maps, note that $\Gamma^{(3)}$ is at $(2\pi, 2\pi)$ of the original Brillouin zone, and so is nominally equivalent to $\Gamma$. However, due to matrix element effects, the $\beta$ and $\varepsilon$ bands are more strongly emphasized at $\Gamma^{(3)}$ than at $\Gamma$ at this photon energy.  

\begin{figure}[t!]
\includegraphics[width=1\columnwidth]{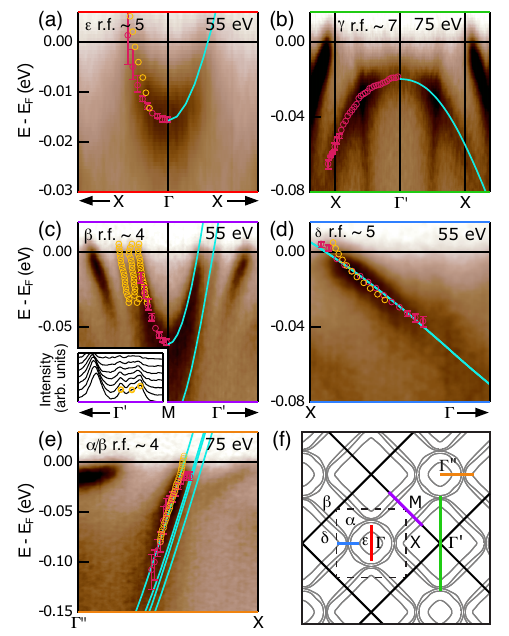}
\caption{Band renormalizations. (a)-(e) Cuts featuring the $\varepsilon,\gamma,\beta,\delta$, and near-degenerate $\alpha/\beta$ bands respectively, as measured at 7~K and with a photon energy of 55~eV for cuts (a),(c)-(d) and 75~eV for cuts (b),(e). EDC- and MDC-derived dispersions are overlaid on the cuts in red and orange respectively. The inset in (c) contains the seven topmost MDCs used in fitting the MDC-derived dispersions in (c), and shows the three-fold splitting of the $\beta$ band (see also Fig. 2(e)).
Error bars reflect the range of peak positions that were attained with different starting fit parameters, while markers indicate the average of those positions. Points without error bars feature ranges smaller than the markers. Theory dispersions (blue) are vertically displaced by a chemical potential shift to align the theory and experiment Fermi momenta, and then divided to match the measured dispersions. The chemical potential shifts used are about 50, -10, -10, 130, and 30~mev for each respective cut. The resulting renormalization factors are approximately 5, 7, 4, 5, and 4 for the $\varepsilon,\gamma,\beta,\delta$, and near-degenerate $\alpha/\beta$ bands respectively. (f) Cut orientations.}
\end{figure}

In the map centered at X (left half of Fig. 1(d)), the $\alpha$, $\beta$, $\delta$, and $\varepsilon$ bands are visible, as are the reconstructed $\delta$ bands, and some splitting of the $\beta$ bands is apparent. Additionally, in the map centered at $\Gamma'$ (right half of Fig. 1(d)), reconstruction of the $\alpha$ and $\beta$ bands is visible. Spectral weight associated with the $\gamma$ band is also present near $\Gamma'$, as the $\gamma$ band lies slightly below the Fermi level for a wide range of momenta and its spectral weight leaks into the Fermi surface map. 

In Fig. 2, we present several high symmetry cuts. In the $\Gamma'-\Gamma-\Gamma'$ cut [Fig. 2(a)], the $\delta,\gamma$, and $\varepsilon$ bands are observed, as are the $\alpha/\beta$ bands, which are nearly degenerate in this direction. The reconstructed $\alpha/\beta$ bands are also observed. In the $\Gamma'-$X$-\Gamma''-$X and the $\Gamma''-\Gamma'-\Gamma''$ cuts [Fig. 2(b),(c)], the matrix elements differ from the $\Gamma'-\Gamma-\Gamma'$ cut, placing greater emphasis on the $\varepsilon$ band and on the band top of the $\gamma$ band, and revealing the reconstructed $\varepsilon$ and $\delta$ bands.  The $\varepsilon$ band unambiguously differentiates the 1313 structure from the 2222 and trilayer structures. The $\delta$ band, which is predicted by LDA calculations to be of predominantly monolayer character (see Fig. 1(e)), serves as additional confirmation of the 1313 structure.

In the M-X-M and $\Gamma'-\Gamma-\Gamma'$ cuts [Fig. 2(d),(e)], an apparent three-fold splitting of the $\beta$ bands is observed. Note that a three-fold splitting is expected from multilayer coupling within the trilayer section of the 1313 structure, but one of those three is accounted for by the $\gamma$ band, so only a two-fold splitting is expected for the $\beta$ band \cite{puphal_unconventional_2023}. A two-fold splitting has also been predicted for the corresponding band in the trilayer structure, although it has not been resolved thus far~\cite{li_fermiology_2017, du2024correlated, zhang_superconductivity_2024}. The observation of the three bands here is therefore inconsistent with the theory predictions. We speculate that the presence of the surface may induce additional symmetry breaking and produce an additional band.

With a detailed understanding of the electronic structure, we quantitatively examine the band dispersions and renormalization effects. In Fig. 3, we compare the calculated and measured dispersions of the $\alpha,\beta,\gamma,\delta$, and $\varepsilon$ bands. The measured dispersions for the bands are derived via the fitting of energy distribution curves (EDCs) and momentum distribution curves (MDCs). We find that the measured dispersions are renormalized by factors of about 4, 4, 7, 5, and 5 to match the calculated dispersions for the $\alpha,\beta,\gamma,\delta$, and $\varepsilon$ bands respectively. Note that according to LDA calculations, the $\alpha$ and $\beta$ bands are primarily of Ni-$d_{x^2-y^2}$ character, while the $\gamma,\delta$ and $\varepsilon$ bands are primarily of Ni-$d_{z^2}$ character, so we find somewhat larger renormalizations in the bands of Ni-$d_{z^2}$ character than in those of Ni-$d_{x^2-y^2}$ character. For comparison, the large renormalizations observed here are on a scale similar to the renormalizations seen in the Fe-based correlated superconductors~\cite{yi_role_2017}.

{\it Important differences between the 1313, 2222, and trilayer La$_4$Ni$_3$O$_{10}$ structures}.\textemdash
In this work we have presented a systematic investigation of the electronic structure in
La$_3$Ni$_2$O$_7$ crystals with the 1313 structure. The main qualitative 
differences between the 1313 structure and the 2222~\cite{yang_orbital-dependent_2024, du2024correlated} and trilayer La$_4$Ni$_3$O$_{10}$~\cite{li_fermiology_2017, du2024correlated, zhang_superconductivity_2024} structures studied previously are the $\varepsilon$ band at $\Gamma$, and the $\delta$ band originating from monolayer Ni orbitals. More quantitatively, we found that in the 1313 structure, the flat portion of the $\gamma$ band approaches about 20~mev below E$_F$, compared to 50~mev below E$_F$ in the 2222 structure~\cite{yang_orbital-dependent_2024}. The $\varepsilon$ and $\gamma$ bands are both of Ni-$d_{z^2}$ orbital character originating in the trilayer, and thus are the bands most directly related to the coupling between Ni-O planes within each trilayer. These differences in the low-lying Ni-$d_{z^2}$ derived bands may be important in ultimately explaining the superconductivity hosted by perovskite nickelates of various layer structures.

We also observed strong renormalizations of the LDA calculated bands relative to the measured bands, with renormalization factors of about 4 for both the Ni-$d_{x^2-y^2}$ derived $\alpha$ and $\beta$ bands and about 5 to 7 for the $d_{z^2}$ derived $\gamma, \delta$, and $\varepsilon$ bands. This result differs from the renormalizations factors of about 2 reported in the 2222~\cite{yang_orbital-dependent_2024} and trilayer La$_4$Ni$_3$O$_{10}$~\cite{li_fermiology_2017} structures for the Ni-$d_{x^2-y^2}$ derived bands, indicating that the different stacking configurations may drastically alter the strength of correlations within different orbitals.

{\it Conclusion}.\textemdash The pressing question in the current quest to understand superconductivity in the perovskite nickelates is: Which phase or phases are responsible for the resistivity drop at high pressures? Given the extensive experimental search in 2222 crystals for a clear signature of a bulk Meissner effect, meaning a signal corresponding to a significant volume fraction of the sample, and the mixed result thus far~\cite{sun_signatures_2023, zhou_evidence_2024, li_pressure-driven_2024}, it seems possible that an alternate structure may host the resistivity drop phenomena. As crystals with the 1313 majority phase also shows a dramatic resistivity drop at similar temperature and pressure regimes~\cite{puphal_unconventional_2023}, a thorough search for the Meissner signal in these crystals is required. Considering that the 1313 structure hosts a $\gamma$ band with a high density of states much closer to E$_F$, we speculate that it may be more conducive to superconductivity.

An alternate scenario may be that the interface between different structures  gives rise to superconductivity. In this case, it would be extremely difficult to detect a Meissner signal if the interface is distributed randomly with a low density. A strategy to investigate this scenario would be to synthesize an engineered heterostructure that contains alternating 2222 and 1313 structures, for example using the method of layer-by-layer molecular beam epitaxy growth.

{\it Acknowledgments}.\textemdash
The work at Stanford University and Stanford Institute for Materials and Energy Sciences is supported by the Department of Energy, Office of Basic Energy Sciences, Division of Materials Sciences and Engineering, under contract DE-AC02-76SF00515 (S.N.A., K.-J.X., E.G.L., B.M, T.P.D., and Z.-X.S.). Use of the Stanford Synchrotron Radiation Lightsource, SLAC National Accelerator Laboratory, is supported by the US Department of Energy, Office of Science, Office of Basic Energy Sciences, under contract no. DE-AC02-76SF00515 (S.N.A., K.-J.X., M.H., D.-H.L., and Z.-X.S.). This research used resources of the Advanced Light Source, which is a DOE Office of Science User Facility under contract no. DE-AC02-05CH11231 (S.N.A., K.-J.X., Y.Z., A.F., S.-K.M., and Z.-X.S.). Computational work was performed on the Sherlock cluster at Stanford University and on resources of the National Energy Research Scientific Computing Center (NERSC), a Department of Energy Office of Science User Facility, using NERSC award BES-ERCAP0027203 (E.G.L., B.M., and T.P.D.).	


\bibliography{bibl_june15_v6.bib}

\end{document}


\title{Supplemental Materials to ``Electronic structure of the alternating monolayer-trilayer phase of La$_3$Ni$_2$O$_7$"}

\author{Sebastien Abadi}
 \altaffiliation[]{These authors contributed equally to this work.}
 \affiliation{%
 Stanford Institute for Materials and Energy Sciences, SLAC National Accelerator Laboratory, 2575 Sand Hill Road, Menlo Park, California 94025, USA
 }%
 \affiliation{%
 Geballe Laboratory for Advanced Materials, Department of Physics and Applied Physics, Stanford University, Stanford, California 94305, USA \\
 }
 \affiliation{%
 Department of Physics, Stanford University, Stanford, CA, 94305, USA}

\author{Ke-Jun Xu}
 \altaffiliation[]{These authors contributed equally to this work.}
 \affiliation{%
 Stanford Institute for Materials and Energy Sciences, SLAC National Accelerator Laboratory, 2575 Sand Hill Road, Menlo Park, California 94025, USA
 }%
 \affiliation{%
 Geballe Laboratory for Advanced Materials, Department of Physics and Applied Physics, Stanford University, Stanford, California 94305, USA \\
 }
 \affiliation{%
 Department of Applied Physics, Stanford University, Stanford, CA, 94305, USA}

\author{Eder G. Lomeli}
 \altaffiliation[]{These authors contributed equally to this work.}
  \affiliation{%
 Stanford Institute for Materials and Energy Sciences, SLAC National Accelerator Laboratory, 2575 Sand Hill Road, Menlo Park, California 94025, USA
 }%
 \affiliation{Department of Materials Science and Engineering, Stanford University, Stanford, CA, 94305, USA}

\author{Pascal Puphal}
  \affiliation{Max Planck Institute for Solid State Research, Heisenbergstraße 1, D-70569 Stuttgart, Germany}

\author{Masahiko Isobe}
 \affiliation{Max Planck Institute for Solid State Research, Heisenbergstraße 1, D-70569 Stuttgart, Germany}

\author{Yong Zhong}
 \affiliation{%
 Stanford Institute for Materials and Energy Sciences, SLAC National Accelerator Laboratory, 2575 Sand Hill Road, Menlo Park, California 94025, USA
 }%
 \affiliation{%
 Geballe Laboratory for Advanced Materials, Department of Physics and Applied Physics, Stanford University, Stanford, California 94305, USA \\
 }
 \affiliation{%
 Department of Applied Physics, Stanford University, Stanford, CA, 94305, USA}

\author{Alexei V. Fedorov}
 \affiliation{Advanced Light Source, Lawrence Berkeley National Laboratory, Berkeley, CA, 94720, USA}

\author{Sung-Kwan Mo}
 \affiliation{Advanced Light Source, Lawrence Berkeley National Laboratory, Berkeley, CA, 94720, USA}

\author{Makoto Hashimoto}
 \affiliation{Stanford Synchrotron Radiation Lightsource, SLAC National Accelerator Laboratory, Menlo Park, CA, 94305, USA}

\author{Dong-Hui Lu}
 \affiliation{Stanford Synchrotron Radiation Lightsource, SLAC National Accelerator Laboratory, Menlo Park, CA, 94305, USA}

\author{Brian Moritz}
 \affiliation{%
 Stanford Institute for Materials and Energy Sciences, SLAC National Accelerator Laboratory, 2575 Sand Hill Road, Menlo Park, California 94025, USA
 }%

\author{Bernhard Keimer}
 \affiliation{Max Planck Institute for Solid State Research, Heisenbergstraße 1, D-70569 Stuttgart, Germany}

\author{Thomas P. Devereaux}
\email{tpd@slac.stanford.edu}
  \affiliation{%
 Stanford Institute for Materials and Energy Sciences, SLAC National Accelerator Laboratory, 2575 Sand Hill Road, Menlo Park, California 94025, USA
 }%
 \affiliation{Department of Materials Science and Engineering, Stanford University, Stanford, CA, 94305, USA}

\author{Matthias Hepting}
\email{hepting@fkf.mpg.de}
 \affiliation{Max Planck Institute for Solid State Research, Heisenbergstraße 1, D-70569 Stuttgart, Germany}

\author{Zhi-Xun Shen}
\email{zxshen@stanford.edu}
 \affiliation{%
 Stanford Institute for Materials and Energy Sciences, SLAC National Accelerator Laboratory, 2575 Sand Hill Road, Menlo Park, California 94025, USA
 }%
 \affiliation{%
 Geballe Laboratory for Advanced Materials, Department of Physics and Applied Physics, Stanford University, Stanford, California 94305, USA \\
 }
 \affiliation{%
 Department of Physics, Stanford University, Stanford, CA, 94305, USA}
  \affiliation{%
 Department of Applied Physics, Stanford University, Stanford, CA, 94305, USA}
 
\date{\today}

\maketitle

\renewcommand{\thefigure}{S\arabic{figure}}

\section{Methods}

\subsection{Crystal growth} Single crystals of the 1313 structure of La$_3$Ni$_2$O$_7$ were grown as described in Puphal, et al. (2023)~\cite{puphal_unconventional_2023}. Details of the structural, electrical, and magnetic characterizations can be found within this reference as well.

\subsection{ARPES measurements} ARPES measurements shown here were performed at Beamline 5-2 at the Stanford Synchrotron Radiation Lightsource, which uses a Scienta DA30L hemispherical electron energy analyzer. The experimentally determined energy resolution was about 15~meV. The beam spot size was approximately 10~µm by 30~µm. The data shown here was taken using photon energies of 55~eV and 75~eV. Samples were cleaved in situ at 10 K or below, and measurements were conducted at 7~K. Chamber pressure was below $3\times  10^{-11}$ Torr throughout the experiment. As the samples are non-superconducting at ambient pressures, we do not expect a gap at the Fermi level. The Fermi level was calibrated by fitting the EDCs at k$_F$ for each cut. Preliminary ARPES measurements were also performed at Beamlines 10.0.1.1 and 10.0.1.2 at the Advanced Light Source, Lawrence Berkeley National Laboratory.

\subsection{Band structure calculations} Density Functional Theory (DFT) calculations for the unbuckled structure were done for band structure and Fermi surface using the Vienna ab initio simulation package (VASP)~\cite{kresse_efficient_1996}, within the projector augmented wave (PAW) formalism~\cite{blochl_projector_1994}. Electronic convergence criteria of $10^{-6}$~eV, and a plane wave cutoff of 600~eV were used with a Gaussian smearing of 0.2~eV. We employed the generalized gradient approximations (GGA) exchange correlation functional of Perdew–Burke-Ernzerhof (PBE)~\cite{perdew_generalized_1996}. A rotationally invariant Hubbard U term was also considered at various energies, but the fully non-interacting approximation was most consistent with experiment~\cite{anisimov_band_1991, dudarev_electron-energy-loss_1998}. Structural relaxations with fixed lattice parameters were done for all compositions until the largest force in the cell was lower than 0.002~eV/Å, with a k-grid of $12\times12\times4$ and $11\times11\times11$ for the unbuckled ($P4/mmm$) and buckled ($Fmmm$) structures respectively.



\subsection{EDC fitting} Example fits of EDCs from which the dispersions in Fig. 3 are derived are shown in Fig. S2-S6. The dispersions shown in the
main text are averaged from the results of several of these fits, with different initial fit parameters.

\begin{figure*}[t!]
\includegraphics[width=1\textwidth]{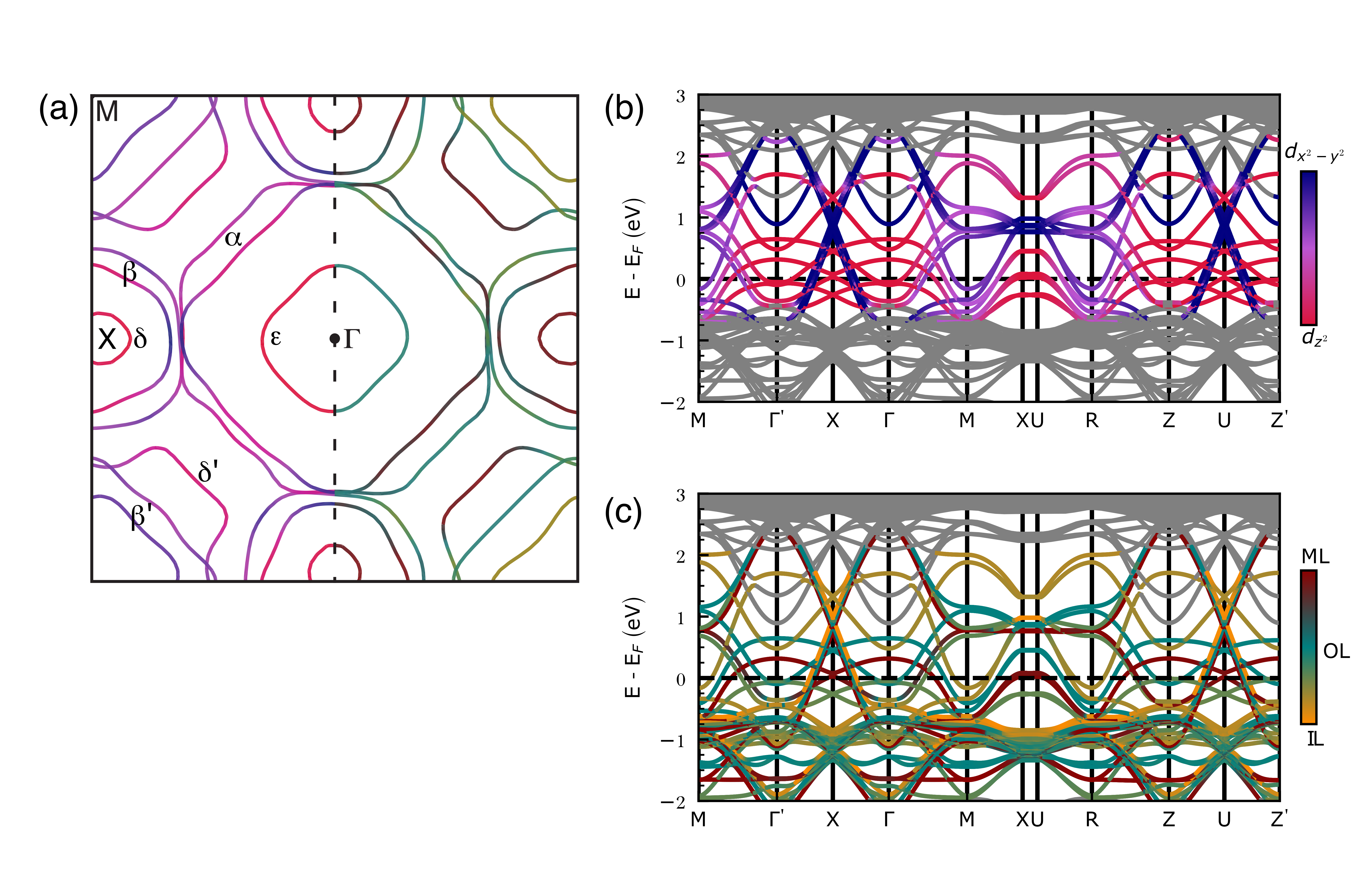}
\caption{ 
(a)-(c) Calculations for the band structure of the 0.7 GPa buckled 1313 structure, with space group $Fmmm$. The color schemes used are the same as in Fig. 1 of the main text. Note that due to Brillouin zone reconstruction, the Brillouin zone in (a) is equivalent to the dashed zone in Fig. 1(b) of the main text. 
Also, due to this reconstruction, the folded $\beta$ and $\delta$ bands are present in (a), and are labelled $\beta'$ and $\delta'$. In (b)-(c), note that $\Gamma$ and $\Gamma'$ are nominally equivalent, but the path $\Gamma'$-$\Gamma$ passes through X.
}
\end{figure*}

\begin{figure*}[t!]
\includegraphics[width=1\textwidth]{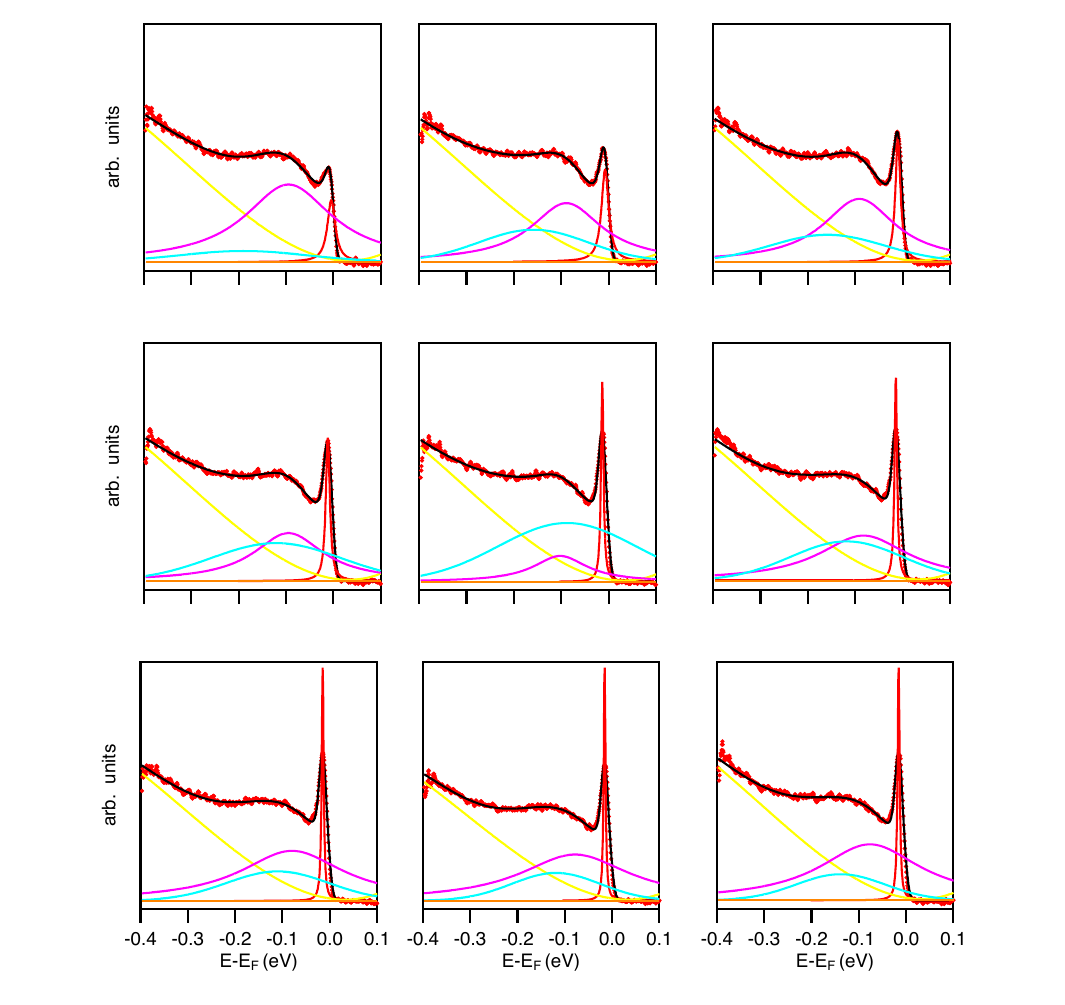}
\caption{ Example EDC fits performed in extracting the dispersion of the $\varepsilon$ band
along the $\Gamma'-\Gamma$ cut, as shown in Fig.  3(a) of the main text. Top left to bottom right here corresponds to left to right in Fig. 3(a).}
\end{figure*}
\newpage
\begin{figure*}[t!]
\includegraphics[width=1\textwidth]{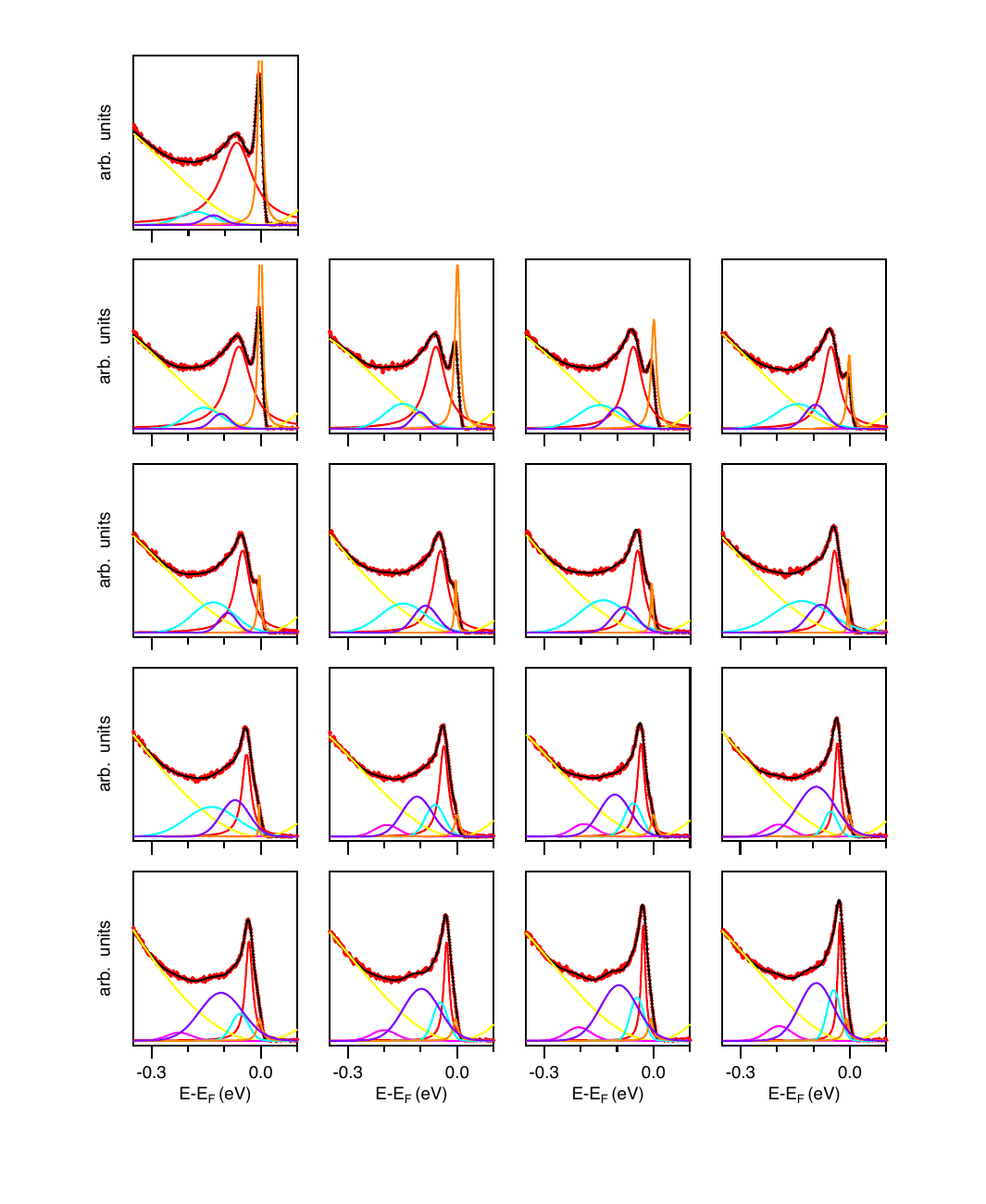}
\caption{ Example EDC fits performed in extracting the dispersion of the $\gamma$ band
along the $\Gamma''-\Gamma'-\Gamma''$ cut, as shown in Fig.  3(b) of the main text. Top left to bottom right here corresponds to left to right in Fig. 3(b). Continued in Fig. S3.}
\end{figure*}
\newpage
\begin{figure*}[t!]
\includegraphics[width=1\textwidth]{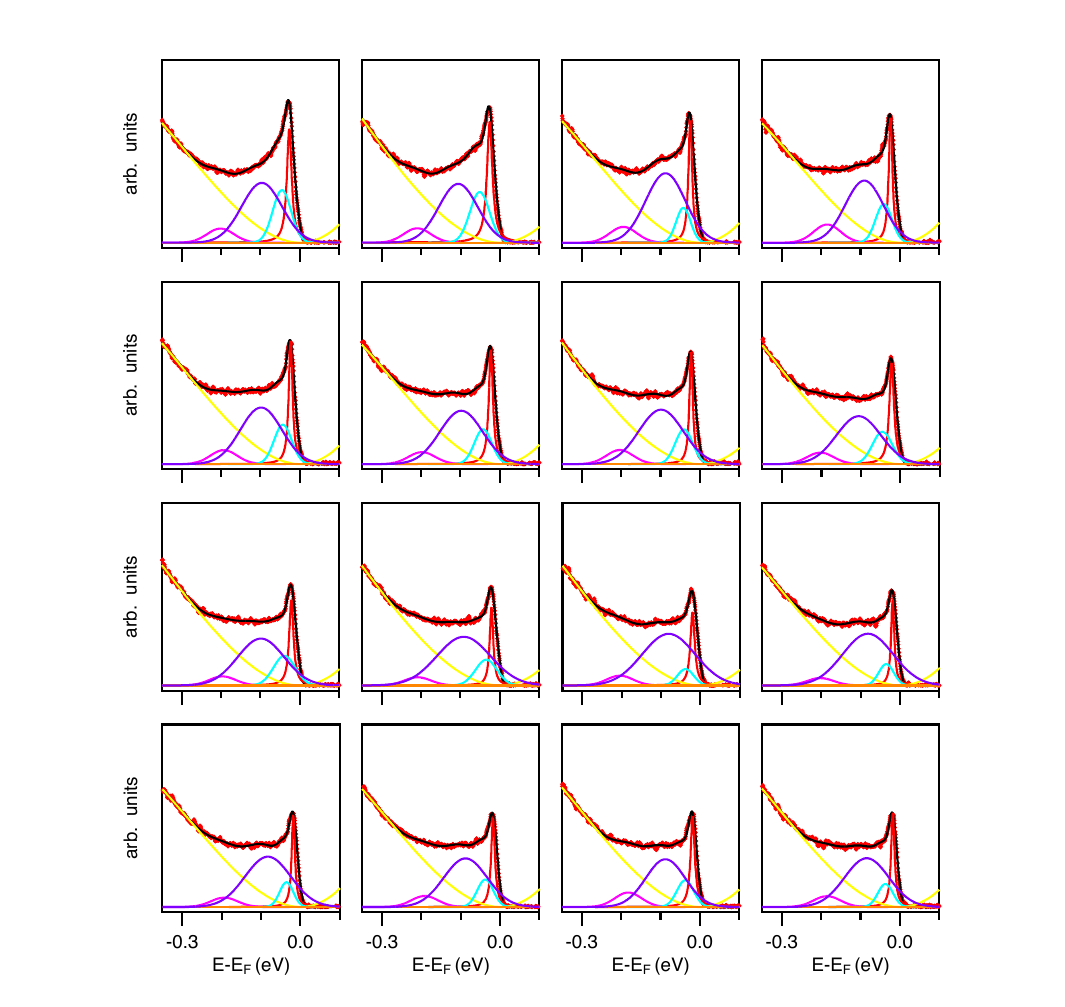}
\caption{EDC fits for the $\gamma$ band continued from Fig. S2.}
\end{figure*}

\begin{figure*}[t!]
\includegraphics[width=1\textwidth]{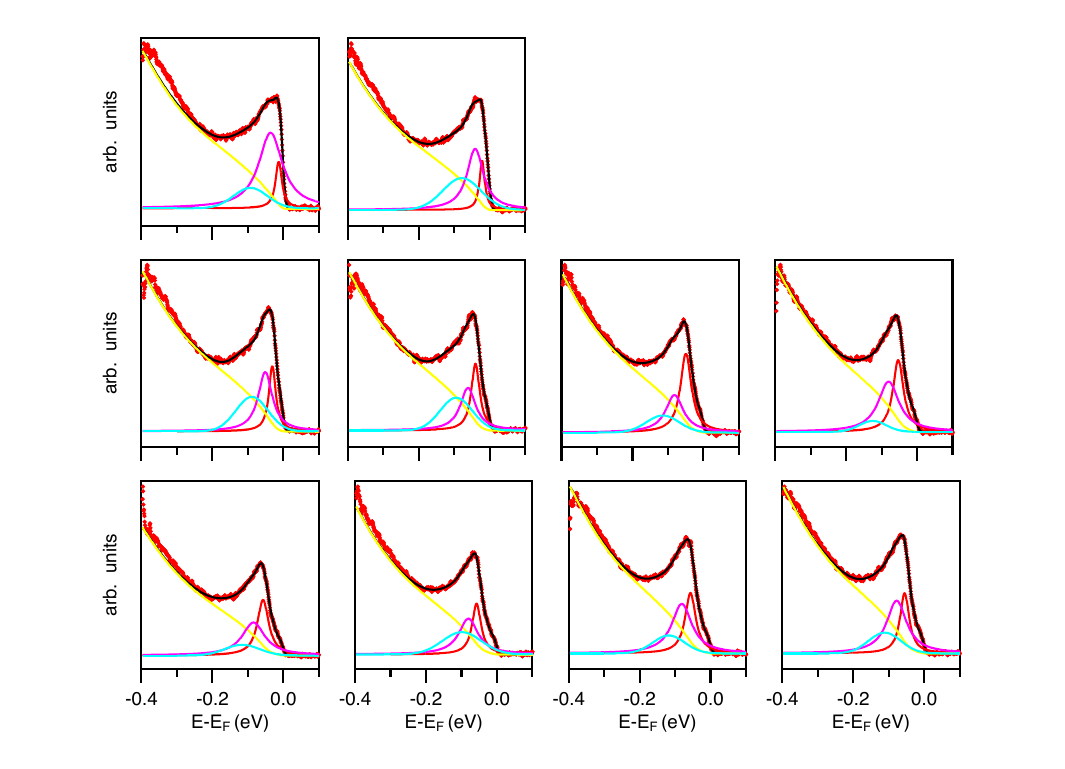}
\caption{ Example EDC fits performed in extracting the dispersion of the $\beta$ band
along the M-X-M cut, as shown in Fig.  3(c) of the main text. Top left to bottom right here corresponds to left to right in Fig. 3(c).}
\end{figure*}

\begin{figure*}[t!]
\includegraphics[width=1\textwidth]{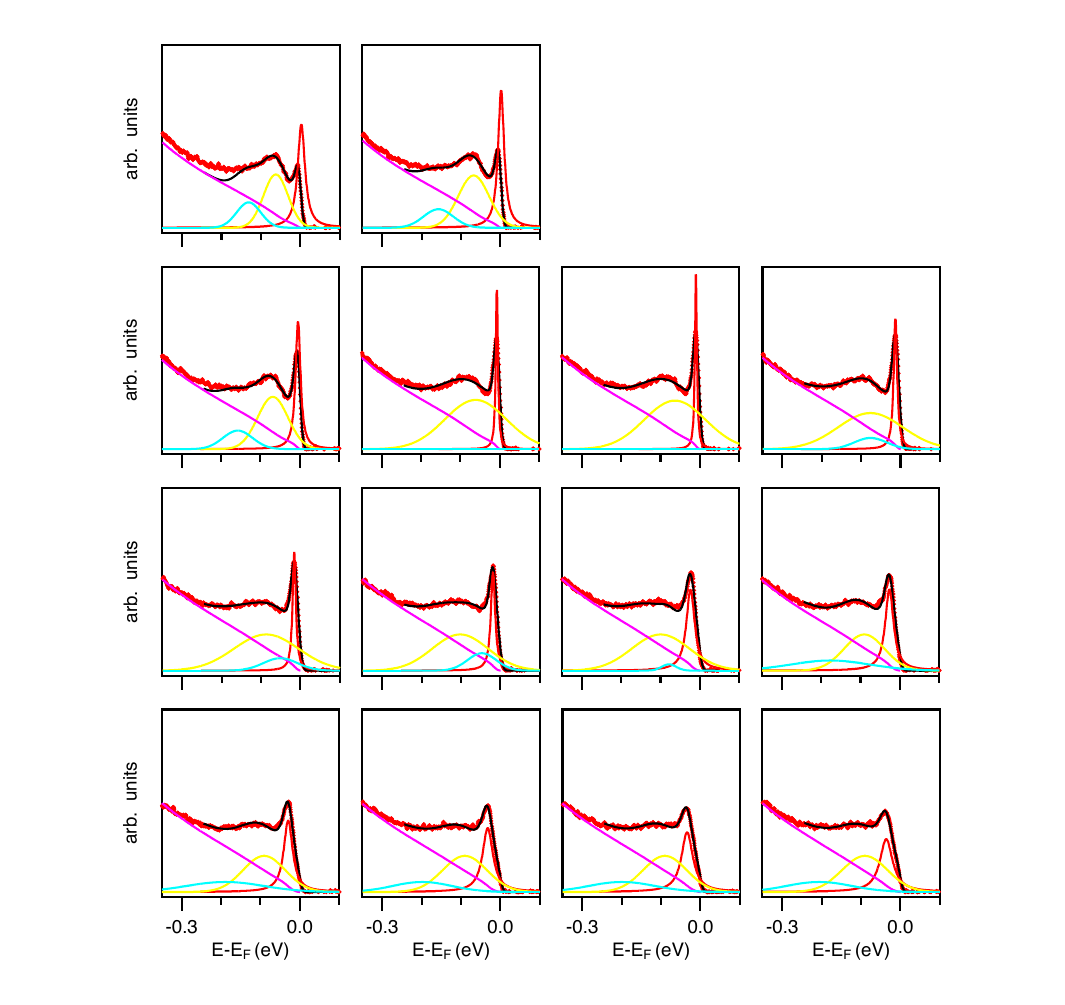}
\caption{ Example EDC fits performed in extracting the dispersion of the $\delta$ band
along the X-$\Gamma$ cut, as shown in Fig.  3(d) of the main text. Top left to bottom right here corresponds to left to right in Fig. 3(d).}
\end{figure*}

\begin{figure*}[t!]
\includegraphics[width=1\textwidth]{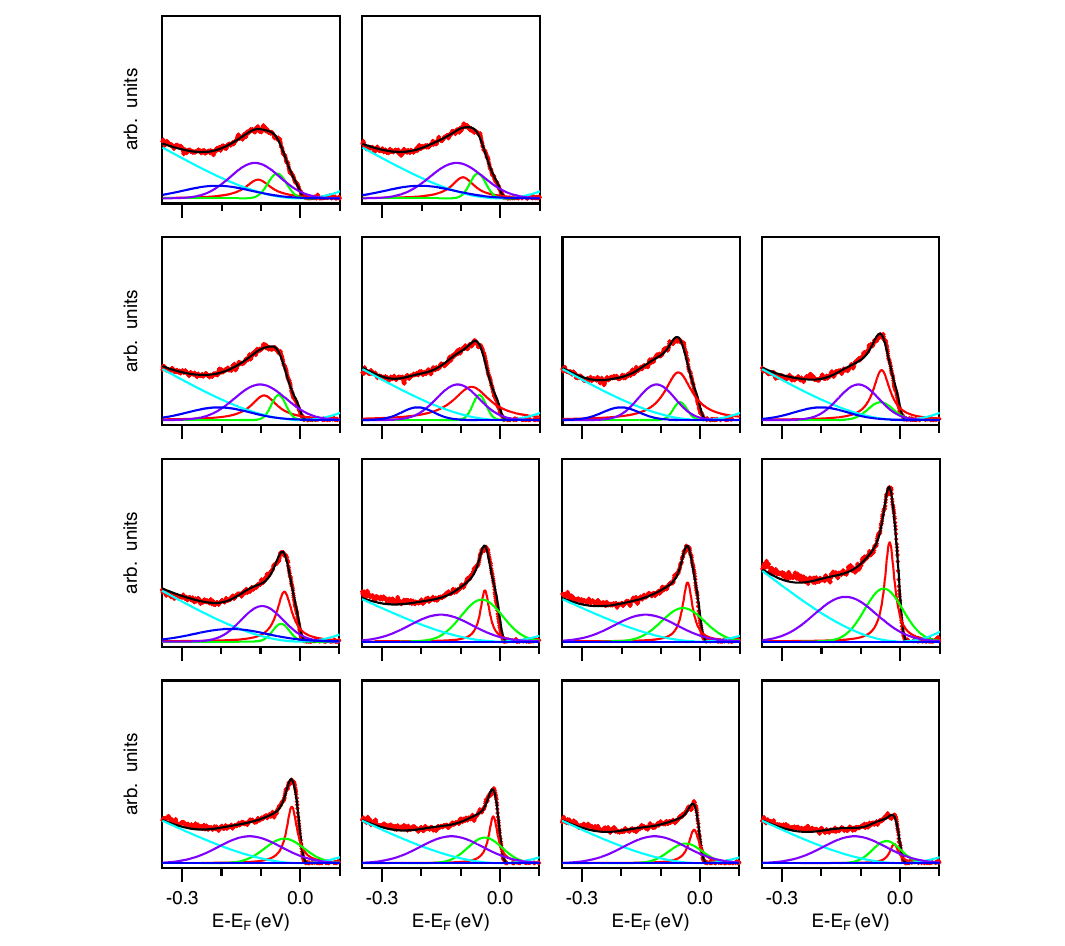}
\caption{ Example EDC fits performed in extracting the dispersion of the nearly degenerate $\alpha/\beta$ bands
along the $\Gamma''-$X cut (see Fig. 3(f)), as shown in Fig.  3(e) of the main text. Top left to bottom right here corresponds to left to right in Fig. 3(e).}
\end{figure*}

\clearpage
\bibliography{bibl_june15_v6_supp.bib}